\documentstyle[twoside,11pt,epsfig,color]{article}
\begin{document}

\title{Scalar Field Collapse with an exponential potential}

\author{Soumya Chakrabarti\footnote{email: adhpagla@iiserkol.ac.in}
 \\
Department of Physical Sciences, \\Indian Institute of Science Education and Research, Kolkata\\ Mohanpur Campus, West Bengal 741252, India.
\date{}
}

\maketitle
\vspace{0.5cm}
{\em PACS Nos. 04.50.Kd; 04.70.Bw
\par Keywords : gravitational collapse, exact solution, spherical symmetry, scalar field, singularity, horizon}
\vspace{0.5cm}

\pagestyle{myheadings}
\newcommand{\be}{\begin{equation}}
\newcommand{\ee}{\end{equation}}
\newcommand{\bea}{\begin{eqnarray}}
\newcommand{\eea}{\end{eqnarray}}

\begin{abstract}
An analogue of the Oppenheimer-Synder collapsing model is treated analytically, where the matter source is a scalar field with an exponential potential. An exact solution is derived followed by matching to a suitable exterior geometry, and an analysis of the visibility of the singularity. In some situations, the collapse indeed leads to a finite time curvature singularity, which is always hidden from the exterior by an apparent horizon.
\end{abstract}

\section{Introduction}
The problem of gravitational collapse of massive stars received significant attention for many years, since the first significant work in this connection by Oppenheimer and Snyder \cite{Opp}. A spacetime singularity is expected to form as an end-state of a continual gravitational collapse and they potentially denote the breakdown of known laws of physics. Though the general speculation is that an event horizon would form to shield this singular state from a distant observer, there are reasonable theoretical examples where no horizon can form and the singularity remains exposed, maybe forever, giving rise to what is known as a naked singularity \cite{Goswami, Joshi}. Therefore it is of paramount importance to investigate what an unhindered gravitational collapse can lead to. For a systematic summarisation of various aspects of gravitational collapse and its possible outcome, we refer to the discussions by Joshi \cite{pankaj1, pankaj2}.                 \\

Gravitational collapse is well studied analytically and numerically, in the framework of General Relativity \cite{Celer, Giambo, Harada, Goswami2, Dadhich, Joshi2}, and modifications of GR as well \cite{Goswami4, scnb1, scnb2, jar, hwang, ghosh}. The exact solution of Einsteins' field equations play a crucial role in understanding a collapsing solution in a detailed manner. The case of a massless scalar field coupled to gravity is of particular interest in gravitational collapse as well as cosmological contexts. One of the major mechanisms for which scalar fields are thought to be responsible is the inflationary scenario \cite{Paddy}. Also, a scalar field with a variety of potential can mimic the evolution of many a kind of matter distribution as discussed by Goncalves and Moss \cite{gong2}. Non-spherical models of scalar field collapse are there in literature \cite{brien, nb}, however, the more popular approach being the spherically symmetric collapse. A number of numerical and analytical work have been done in recent years on spherically symmetric massless scalar field collapse models from the perspective of cosmic censorship hypothesis \cite{goswami, giambo}. A zero mass scalar field collapse and the possibility of the formation of a naked singularity was discussed in a series of work by Christodoulou \cite{christo1, christo2, christo3}. With the help of a numerical analysis, Goldwirth and Piran, showed that a scalar field collapse leads to a singularity which is cut-off from the exterior observer by an event horizon \cite{piran}. Investigations on collapsing gravitational systems with a scalar field revealed that black hole threshold for classes of massless scalar fields can show universality and power-law scaling of the black hole mass, which correspond to a critical phenomena \cite{chop, brady, gund, Gundlach1}, and constitutes a potentially rich store of possible investigations.     \\

The collapse scenario of massive scalar fields have also been studied \cite{giambo, gong1, gong2, goswami2, koyel, cai1, cai2}. A scalar field collapse, along with a negative cosmological constant, can lead to the formation of a naked singularity as discussed by Baier, Nishimura and Stricker \cite{baier} very recently. A conformally flat massive scalar field collapse with spatial homogeneity was discussed by Chakrabarti and Banerjee \cite{scnb3}. Such a scalar field collapse with a power-law potential, without any apriori choice of equation of state, can lead to the formation of a central singularity which is cutoff from the exterior by an apparent horizon. The additional relevance of such investigations stems from the present importance of a scalar field as the dark energy \cite{varun, sami}, the agent responsible for the late time acceleration of the universe whose distribution remains an open aspect of investigation.               \\

However, the number of known exact solutions for self-gravitating scalar field collapse is rather limited. The aim of the present work is to look at the collapse of a massive scalar field. A common functional form for the self-interaction potential is an exponential dependence upon the scalar field. A homogeneous isotropic cosmological model driven by a scalar field with an exponential potential was studied and a solution with power-law inflation was shown to be an attractor by Halliwell \cite{hall}. An exponential potential is predicted to be found in higher-order \cite{whitt} or higher-dimensional gravity theories \cite{green}. The cosmological nature of the universe filled with a scalar field, with an exponential potential, has been studied for both homogeneous and inhomogeneous scalar fields \cite{barrow}. It is always a challenge to write the highly non-linear field equations in an integrable form. The exact time-evolution here is studied analytically for a homogeneous scalar field in a flat FLRW spacetime. In a recent approach by Harko, Lobo and Mak \cite{harko}, a new formalism for the analysis of scalar fields in flat isotropic and homogeneous cosmological models was presented. The basic evolution equations of the model were reduced to a first order non-linear differential equation. The transformation introduced therein, is used to simplify the evolution equations here.          \\

In section $(2)$ we define the basic equations and formulate the problem by introducing the transformations. Section $(3)$ introduces the exponential interaction of the scalar field and the exact time-evolutions for the scale factor, scalar field and the curvature scalars. In secton $(4)$ a smooth boundary matching with an exterior Vaidya metric is discussed in brief and section $(5)$ includes an investigation of the visibility of the ultimate spacetime singularity. The final section briefs the results obtained and gives a conclusion.          \\

\section{Mathematical formulaion}
We write the metric for a spherically symmetric spacetime as 

\begin{equation}
\label{metricltb}
ds^2=dt^2-T(t)^2(dr^2+r^2d\Omega^2).
\end{equation}

The time evolution is governed solely by the function $T(t)$. This indeed is a simple case, but this would lead to some tractable solutions so that the possibility of collapse can be investigated.                             \\
When a scalar field $\phi$ is minimally coupled to gravity, the relevant action is given by 
\begin{equation}\label{action}
\textit{A}=\int{\sqrt{-g}d^4x[R+\frac{1}{2}\phi^\mu\phi_\mu-V(\phi) + L_{m}]},
\end{equation}

where $V(\phi)$ is the scalar potential and $L_{m}$ is the Lagrangian density for the fluid distribution. In this particular case, we assume that there is no fluid contribution in the action, thus, $L_{m}=0$.

\par From this action, the contribution to the energy-momentum tensor from the scalar field $\phi$ can be  written as
\begin{equation}\label{minimallyscalar}
T^\phi_{\mu\nu}=\partial_\mu\phi\partial_\nu\phi-g_{\mu\nu}\Bigg[\frac{1}{2}g^{\alpha\beta}\partial_\alpha\phi\partial_\beta\phi-V(\phi)\Bigg]. 
\end{equation}

We assume the scalar field to be spatially homogeneous, i.e., $\phi=\phi(t)$. With this assumption, the Einstein field equations for the metric (\ref{metricltb}) can be written as (in the units where $8 \pi G = 1$)
\begin{equation} \label{fe1ltb}
3\Bigg(\frac{\dot{T}}{T}\Bigg)^{2} = \frac{\dot{\phi}^{2}}{2}+V\left( \phi \right),
\end{equation}  
\begin{equation} \label{fe2ltb}
-2\frac{\ddot{T}}{T}-\Bigg(\frac{\dot{T}}{T}\Bigg)^{2} = \frac{\dot{\phi}^{2}}{2}-V\left( \phi \right).
\end{equation}
The evolution equation for the scalar field is given by
\begin{equation} \label{philtb}
\ddot{\phi}+3\frac{\dot{T}}{T}\dot{\phi}+\frac{dV(\phi)}{d\phi} = 0.  
\end{equation}                                    \\

The overhead dot denotes the derivative with respect to the time-coordinate $t$. We will restrict our study to collapsing models, which satisfy the condition that radius of the two-sphere is a monotonically decreasing function of time. Therefore we discuss only those cases where $\frac{\dot{T}}{T}<0$ is satisfied.        \\

By substituting $\frac{\dot{T}}{T}$ from Eq. (\ref{fe1ltb}) into Eq. (\ref{philtb}), one obtains the basic equation describing the scalar field evolution as
\begin{equation} \label{philtb1}
\ddot{\phi}-\sqrt{3}\dot{\phi}\sqrt{\frac{\dot{\phi}^{2}}{2}+V\left(\phi \right)}+\frac{dV}{d\phi }=0.  
\end{equation}

Defining a new function $f(\phi)$ so that $\dot{\phi}^{2}=f(\phi)$, and changing the independent variable from $t$ to $\phi$, eq. (\ref{philtb1}) can be written as
\begin{equation} \label{philtb2}
\frac{1}{2}\frac{df(\phi)}{d\phi}-\sqrt{3}\sqrt{\frac{f(\phi)}{2}+V(\phi)}\sqrt{f(\phi)}+\frac{dV}{d\phi}=0,  
\end{equation}
which may be reorganized into the following form
\begin{equation} \label{philtb3}
\frac{\frac{1}{2}\frac{df(\phi)}{d\phi}+\frac{dV}{d\phi}}{2\sqrt{\frac{f(\phi)}{2}+V(\phi)}}-\frac{\sqrt{3}}{2}\sqrt{f(\phi)}=0.  
\end{equation}                        \\

The transformations introduced by Harko et. al. is considered thereafter, (for step by step systematic treatment, we refer to \cite{harko}), defined by $F(\phi)=\sqrt{\frac{f(\phi)}{2}+V(\phi)}$, $F(\phi)=u(\phi)\sqrt{V(\phi)}$ and $u(\phi)=\cosh G(\phi)$; such that one can simplify (\ref{philtb3}) enough to arrive at the basic equation governing the dynamics of the scalar field collapse given as
\begin{equation} \label{finltb}
\frac{dG}{d\phi}+\frac{1}{2V}\frac{dV}{d\phi}\coth G-\sqrt{\frac{3}{2}}=0.
\end{equation}

For a flat FRW spacetime, it can be shown that the functions $f(\phi)$ and $F(\phi)$ are related to the Hubble function and its time derivative using the field equations (\ref{fe1ltb}) and (\ref{fe2ltb}). Another similar approach was considered in \cite{bond} where the Hubble function was assumed to be a function of the scalar field $\phi$. The time evolution of the two-sphere defined by $T(t)$ can be written in terms of the scalar field, by the equation
\begin{equation}  \label{Tt}
\frac{1}{T(\phi)}\frac{dT(\phi)}{d\phi}=-\frac{1}{\sqrt{6}}\coth{G(\phi)}.
\end{equation}                        
The fact that $\frac{\dot{T(t)}}{T(t)}<0$ for a collapsing geometry is considered here.

For a large number of choices of the functional form of the self-interacting potential, the first order evolution equation, Eq.(\ref{finltb}) can be solved exactly or parametrically and a collapsing model can be discussed extending the discussion, provided they match the junction conditions. Here we only present a special case, a simple example of an exponential potential, such that a complete collapsing scenario can be investigated. Relevant possible choices of the potential and their dynamical scenario in case of scalar field cosmologies are discussed in \cite{harko}.

\section{Exact solution}
If $V^{\prime}/V=\sqrt{6}\alpha_{0}=$ constant, the scalar field self-interaction potential is of the
exponential form,
\begin{equation} \label{potential}
V=V_{0}e^{\left(\sqrt{6}\alpha_{0}\phi\right)}.
\end{equation}

Taking into account Eq. (\ref{potential}), Eq. (\ref{finltb}) takes the form
\begin{equation} \label{evoexp}
\frac{dG}{d\phi}+\sqrt{\frac{3}{2}}\left(\alpha_{0}\coth G-1\right)=0.
\end{equation}

A particular solution of the field equations corresponds to the case $G(\phi)=G_{0}=\mathrm{constant}$. In this case Eq.(\ref{evoexp}) is identically satisfied, with $G_{0}$ given by
\begin{equation}
G_{0}=\mathrm{arccoth}\left(\frac{1}{\alpha_{0}}\right).
\end{equation}

From Eq.(\ref{Tt}) it follows that the scale factor can be obtained as a function of the scalar field as
\begin{equation} \label{evoscale}
T(\phi)=T_{0}e^{-{\phi}/{\sqrt{6}\alpha_{0}}},  
\end{equation}
where $T_{0}$ is a constant of integration. The time variation of the scalar field is determined from the relation $G(\phi)=\mathrm{arccosh} \sqrt{1+\frac{\dot{\phi}^{2}}{2V(\phi)}}$ as
\begin{equation} \label{phidot}
\dot{\phi}=\pm \sqrt{2V_{0}}\Bigg(\frac{\alpha_{0}}{({\alpha_{0}}^2+1)^{\frac{1}{2}}}\Bigg) e^{\sqrt{3/2}\alpha_{0}\phi}.
\end{equation}
It is straightforward to integrate (\ref{phidot}) to write the exact evolution of scalar field with respect to time as
\begin{equation}
e^{-\sqrt{3/2}\alpha_{0}\phi}=\mp \sqrt{3V_{0}}\Bigg(\frac{\alpha_{0}^2}{({\alpha_{0}}^2+1)^{\frac{1}{2}}}\Bigg)(t-t_{0}),
\end{equation}
where $t_{0}$ is an arbitrary constant of integration.
With the help of Eq. (\ref{evoscale}), one can obtain the exact time evolution of the collapsing scalar field in the form
\begin{equation}
T(t)=T_{0}\Bigg[\mp \sqrt{3V_{0}}\Bigg(\frac{\alpha_{0}^2}{({\alpha_{0}}^2+1)^{\frac{1}{2}}}\Bigg)(t-t_0)\Bigg]^{\frac{1}{3\alpha_{0}^{2}}}.
\end{equation}                
Since for a collapsing scenario, $\dot{T(t)}<0$, it is easy to check that one must choose the negative signature inside the parenthesis to write the time-evolution, thus giving the time evolution as
\begin{equation}\label{exactevo}
T(t)=T_{0}[N_{0}(t_0-t)]^{\frac{1}{3{\alpha_{0}}^2}}.
\end{equation}           

We have written $N_{0}=\sqrt{3V_{0}}\Big(\frac{\alpha_{0}^2}{({\alpha_{0}}^2+1)^{\frac{1}{2}}}\Big)$, which must always be greater than zero. It is straightforward to note that radius of the two-sphere $r T(t)$ goes to zero when $t \rightarrow t_{0}$, giving rise to a finite time zero proper volume singularity.       \\

\begin{figure}[h]
\begin{center}
\includegraphics[width=0.4\textwidth]{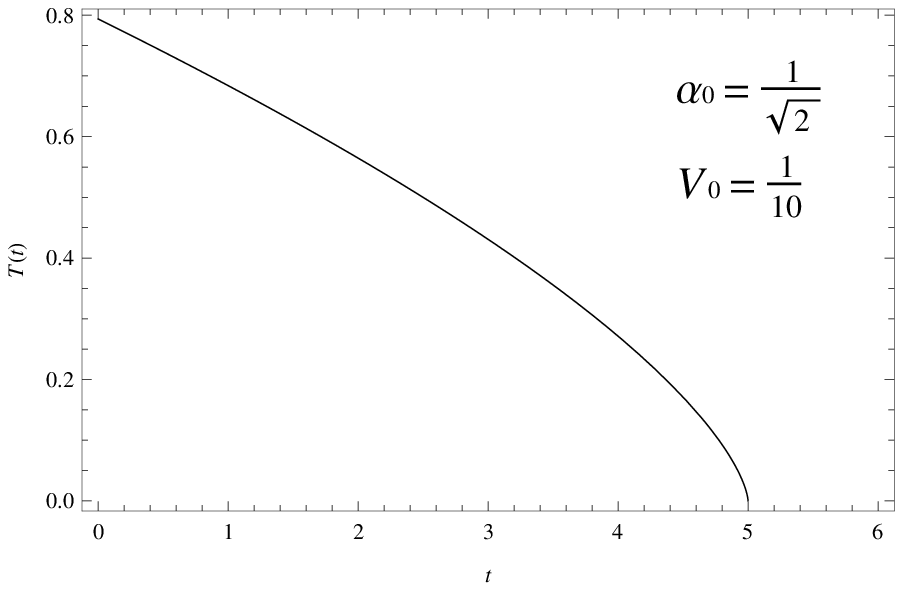}
\includegraphics[width=0.4\textwidth]{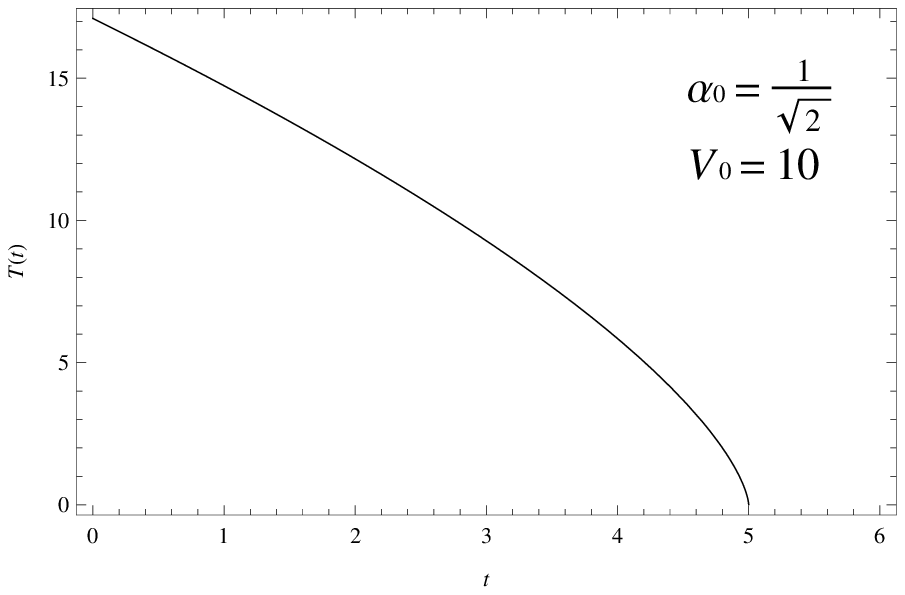}
\caption{\bf Time-evolution of the function $T(t)$ for different values of $V_{0}$}
\end{center}
\label{fig:ltb}
\end{figure}

Since, $N_{0}$ must always be real and positive, it is obvious that one must always choose positive values of $V_{0}$. We plot the evolution of $T(t)$ with respect to $t$ in Figure $(1)$ for different values of $V_{0}$. Without any loss of generality, one can assume $T_{0}=1$. The choice of $\alpha_{0}$ is restricted by the constraints developed from the boundary matching discussed in the next section. The spherical body, collapses with time almost uniformly until it reaches $t=t_{0}=t_{s}$, where it hurries towards a zero proper volume singularity. This behavior is not affected by different choices of $V_{0}$, only the scale of the time-interval changes. We have presented here two specific examples with $V_{0}=10$ and $V_{0}=\frac{1}{10}$.        \\

One must look into the behavior of Ricci and Kretschmann curvature scalars to judge the nature of spacetime singularity. The curvature scalars can be written from the metric (\ref{metricltb}) as

\begin{equation}\label{ricci}
R=-6\Bigg[\frac{\ddot{T(t)}}{T(t)}+\frac{\dot{T(t)}^2}{T(t)^2}\Bigg],
\end{equation}
and
\begin{equation}\label{kret}
K=6\Bigg[\frac{\ddot{T(t)}^2}{T(t)^2}+\frac{\dot{T(t)}^4}{T(t)^4}\Bigg].
\end{equation}

Using equation (\ref{exactevo}), the expressions for the scalars can be simplified into
\begin{equation}
R=\frac{2}{{\alpha_{0}}^2}\Bigg(1-\frac{2}{3{\alpha_{0}}^2}\Bigg)\frac{1}{(t-t_0)^2},
\end{equation}
and
\begin{equation}
K=\frac{2}{3{\alpha_{0}}^4}\Bigg[\frac{1}{9{\alpha_{0}}^4}+\Bigg(\frac{1}{3{\alpha_{0}}^2}-1\Bigg)^2\Bigg]\frac{1}{(t-t_0)^4}.
\end{equation}

One can clearly see, for all values of $\alpha_{0}$ $({\alpha_{0}}^2 \neq \frac{2}{3})$, both ricci and kretschmann scalar diverges to infinity when $t \rightarrow t_{0}$. Therefore the collapsing sphere discussed here ends up in a curvature singularity.                         \\

Singularities formed in collapse can be shell focussing or shell crossing in nature. For a spherically symmetric collapse the shell focusing singularity occurs at $g_{\theta\theta} = 0$. It is evident from (\ref{metricltb}) and (\ref{exactevo}) that $g_{\theta\theta} = rT(t) \rightarrow 0$ when $t \rightarrow t_{0}$. Thus the curvature singularity indeed is a shell-focussing one in nature.

\section{Matching of the interior space-time with an exterior geometry}
For a complete and consistent analysis of gravitational collapse, proper junction conditions are to be examined carefully which allow a smooth matching of an exterior geometry with the collapsing interior. Any astrophysical object is immersed in vacuum or almost vacuum spacetime, and hence the exterior spacetime around a spherically symmetric star is well described by the Schwarzschild geometry. Moreover it was extensively shown by Goncalves and Moss \cite{gong2} that any sufficiently massive collapsing scalar field can be formally treated as collapsing inhomogeneous dust. From the continuity of the first and second differential forms, the matching of the sphere to a Schwarzschild spacetime on the boundary surface, $\Sigma$, is extensively worked out in literature \cite{santos, chan, Kolla, Maharaj}.         \\

However, conceptually this may lead to an inconsistency since the treatment allowed for a dust collapse may not be valid for a scalar field in general. For instance, since Schwarzschild has zero scalar field, such a matching would lead to a discontinuity in the scalar field, which means a delta function in the gradient of
the scalar field. As a consequence, there will appear square of a delta function in the stress-energy, which is definitely an inconsistency. In modified theories of gravity an alternative scenario is treated sometimes where the exterior is non-static, however, the solar system experiments constrain heavily such a scenario. Another possible way to avoid such a scenario can perhaps be allowing jump in the curvature terms in the field equations, however, this must result in surface stress energy terms, which on realistic collapsing models must have observational signatures and can be established via experimental evidences \cite{Goswami4}.        \\

Following references \cite{pankajritu, pankajritu2, koyel}, we match the spherical ball of collapsing scalar field to a Vaidya exterior across a boundary hypersurface defined by $\Sigma$. The metric just inside $\Sigma$ is,
\begin{equation}\label{interior}
d{s_-}^2=dt^2-T(t)^2dr^2-r^2 T(t)^2d{\Omega}^2,
\end{equation}
and the metric in the exterior of $\Sigma$ is given by
\begin{equation}\label{exterior}
d{s_+}^2=(1-\frac{2M(r_v,v)}{r_v})dv^2+2dvdr_v-{r_v}^2d{\Omega}^2.
\end{equation}
Matching the first fundamental form on the hypersurface we get
\begin{equation}\label{cond1}
{\frac{dv}{dt}}_{\Sigma}=\frac{1}{\sqrt{1-\frac{2M(r_v,v)}{r_v}+\frac{2dr_v}{dv}}}
\end{equation}
and
\begin{equation}\label{cond2}
r_v = r T(t) = rT_{0}[N_{0}(t_0-t)]^{\frac{1}{3{\alpha_{0}}^2}}.
\end{equation}
Matching the second fundamental form yields,
\begin{equation}\label{cond3}
rT(t)=r_v\left(\frac{1-\frac{2M(r_v,v)}{r_v}+\frac{dr_v}{dv}}{\sqrt{1-\frac{2M(r_v,v)}{r_v}+\frac{2dr_v}{dv}}}\right)
\end{equation}
Using equations (\ref{cond1}), (\ref{cond2}) and (\ref{cond3}) one can write
\begin{equation}\label{dvdt2}
\frac{dv}{dt}=\frac{T^2-\frac{r}{3}}{T^2-\frac{2M}{r}T}.
\end{equation}
From equation (\ref{cond3}) one obtains
\begin{equation}\label{M}
M=\frac{r^{-1}T^{-1}+\frac{r}{9}T^{-5}+\sqrt{\frac{1}{r^{2}}T^{-2}+\frac{r^2}{81}T^{-10}-\frac{2}{9}T^{-6}}}{\frac{4}{r^2}T^{-2}}.
\end{equation}
Matching the second fundamental form we can also write the derivative of $M(v,r_v)$ as
\begin{equation}\label{dM}
M{(r_v,v)}_{,r_v}=\frac{M}{rT}-\frac{2r^2}{9T^{4}}.
\end{equation}
Equations (\ref{cond2}), (\ref{dvdt2}), (\ref{M}) and (\ref{dM}) completely specify the matching conditions at the boundary of the collapsing scalar field.

\section{Visibility of singularity}
Whether the ultimate spacetime singularity is visible to an exterior observer depends on the formation of an apparent horizon. Such a surface is defined as
\begin{equation}
g^{\mu\nu}Y_{,\mu}Y_{,\nu}=0,
\end{equation}    
where $Y(r,t)$ is the proper radius of the collapsing sphere. Using (\ref{metricltb}) and (\ref{exactevo}), one can express this equation as
\begin{equation}
N_{0}(t_{0}-t_{ap})=\Bigg(\frac{9{{\alpha_{0}}^2}\delta}{{T_{0}}^2{N_{0}}^2}\Bigg)^{\frac{3{\alpha_{0}}^2}{2-6{\alpha_{0}}^2}}.
\end{equation}
Here, $\delta$ is a constant of separation. Thus the time formation of apparent horizon can be expressed as
\begin{equation}\label{apparent}
t_{ap}=t_{0}-\frac{1}{N_{0}}\Bigg(\frac{9{\alpha_{0}}^2\delta}{{T_{0}}^2{N_{0}}^2}\Bigg)^{\frac{3{\alpha_{0}}^2}{2-6{\alpha_{0}}^2}}.
\end{equation}
Both $\alpha_{0}$ and $N_{0}$ depend on the choice of potential given their definitions $V=V_{0}e^{\left(\sqrt{6}\alpha_{0}\phi\right)}$ and $N_{0}=\sqrt{3V_{0}}\Big(\frac{\alpha_{0}^2}{({\alpha_{0}}^2+1)^{\frac{1}{2}}}\Big)$. The arbitrary constants $T_{0}$ and $\delta$ can be estimated from proper matching of the collapsing scalar field with an appropriate exterior geometry and on initial conditions.          \\
From (\ref{exactevo}), one can write at $t=t_{ap}$,
\begin{equation}\label{tdot}
\frac{\dot{T}}{T}=\frac{1}{3{\alpha_0}^2(t_{ap}-t_0)} < 0.
\end{equation}
Since we are dealing with a geometry where the sphere must always decrease in volume with respect to time, this expression is consistent iff $t_{0}>t_{ap}$. This means that the apparent horizon, if any, must always form before the formation of singularity.        \\
To determine the collapse outcome, one needs to find if nonspacelike trajectories emanate from the singularity and reach an observer. The singularity is at least locally naked if there are future directed nonspacelike curves that reach faraway observers. In the present case the time of formation of singularity $t_{0}$ is independent of $r$ and therefore is not a central singularity. The entire collapsing body reaches the singularity simulteneously at $t = t_{0}$. This kind of singularity is always expected to be covered by the formation of an apparent horizon as already discussed by Joshi, Goswami and Dadhich \cite{naresh}. The result that apparent horizon must always form before the formation of singularity, is therefore a consistent result.

\section{Discussion}
Studies of exact solutions and their properties, symmetries, local geometries and singularities - play a non-trivial role in general rel­ativity \cite{mac1, mac2}, even in the current context. Finding non-trivial solutions to the Einstein equations requires some reduction of the problem, which usually is done by exploiting symmetries or other properties. As a result, there is no single method preferred for finding solutions to the Einstein equations. Although many exact solutions exist, very few pose physically interesting results. From this point of view, investigations regarding a gravitational collapse when a scalar field is minimally coupled to gravity are already there in literature, but usually consisting of a massless scalar field. There are only very limited amount of work on massive scalar field collapse and that too in a very restricted scenario. We have discussed a spherically symmetric collapse of a massive scalar field where an exponential potential describes the self-interaction. The scalar field is chosen to be spatially homogeneous, representing the scalar field analogue of Oppenheimer-Snyder collapse \cite{Opp}.              

\par A systematic discussion is presented, assuming the matter contribution $L_{m}$ to be zero. Following the approach introduced by Harko et al \cite{harko} in a cosmological context, the Klein-Gordon equation describing the dynamics of the scalar field is simplified considerably into a first order non-linear differential equation. A simple exact solution predicts the end state of the collapse to be a finite time shell-focusing singularity. The evolution of the system is found to be independent of different parameters defining the self-interacting potential $V=V_{0}e^{\left(\sqrt{6}\alpha_{0}\phi\right)}$. A proper boundary matching with an exterior Vaidya geometry is discussed ensuring the consistency of the geometry. The visibility of the end-state is sensitive over different choices of initial conditions, as discussed in section $(5)$. The collapse is simulteneous and results in a singularity which acts as a sink for all the curves of the collapsing congruence, and the volume elements shrink to zero along all the collapsing trajectories. An apparent horizon is always expected to form before the formation of zero proper volume singularity, which therefore remains hidden forever.
\par The formalism presented here is indeed a very simple case, but can perhaps be interesting for pedagogical purposes. The reduction of the klein-gordon equation into a first order differential equation can be fairly useful in many cosmological contexts. Exact and approximate solutions of massive scalar field collapse is an interesting subject and using this formalism the study can be expanded, for example for hyperbolic cosine potentials. Approximate solutions can be constructed as well in the limiting cases of the scalar field kinetic energy and potential energy dominance.

\section*{Acknowledgement}
The author thanks Professor Narayan Banerjee for invaluable insights and helpful discussions.

\end{document}